\documentstyle[pra,aps,epsf]{revtex}

\begin{document}

\draft

\twocolumn[\hsize\textwidth\columnwidth\hsize\csname@twocolumnfalse%
\endcsname

\title{Anomalous diffusion in disordered media and random quantum spin chains}

\author{Ferenc Igl\'oi$^{1,2}$ and Heiko Rieger$^{3,4}$}

\address{
$^1$ Research Institute for Solid State Physics, 
H-1525 Budapest, P.O.Box 49, Hungary\\
$^2$ Institute for Theoretical Physics,
Szeged University, H-6720 Szeged, Hungary\\
$^3$ Institut f\"ur Theoretische Physik, Universit\"at zu K\"oln, 
     50923 K\"oln, Germany\\
$^4$ HLRZ, Forschungszentrum J\"ulich, 52425 J\"ulich, Germany\\
}

\date{February 27, 1998}

\maketitle

\begin{abstract}
Using exact expressions for the persistence probability and for the leading
eigenvalue of the Focker-Planck operator of a random walk in a random
environment we establish a fundamental relation between the statistical
properties of anomalous diffusion and the critical and off-critical
behavior of random quantum spin chains. Many new exact results are obtained
from this correspondence including the space and time correlations of
surviving random walks and the distribution of the gaps of the corresponding
Focker-Planck operator. In turn we derive analytically the dynamical
exponent of the random transverse-field Ising spin chain in the
Griffiths-McCoy region.
\end{abstract}


]

\newcommand{\bc}{\begin{center}}
\newcommand{\ec}{\end{center}}
\newcommand{\be}{\begin{equation}}
\newcommand{\ee}{\end{equation}}
\newcommand{\beqn}{\begin{eqnarray}}
\newcommand{\eeqn}{\end{eqnarray}}

Ultraslow dynamics is a common feature of low-dimensional systems with
quenched disorder in particular in the vicinity of a critical
point. One of the well known examples in this respect is the
one-dimensional diffusion process in a random media, when - in the
absence of an average drift $\delta_{RW}$ - the mean-square
displacement grows very slowly like \cite{sinai} 
\be
[\langle X^2(t)\rangle ]_{\rm av}\sim\ln^4t
\label{sinai}
\ee
in contrast to the normal diffusive behavior $\langle
X^2(t)\rangle\sim t$ in the homogeneous case. The 
diffusion process remains anomalous for sufficiently small
average drifts $0<\delta_{RW}<\delta_{RW}^+$, when the average
displacement has an algebraic time dependence:
\be
[\langle X(t)\rangle ]_{\rm av} \sim t^{\mu}\;,
\label{tmu}
\ee
where the exponent $0<\mu=\mu(\delta_{RW}) \le 1$ is a continuous
function of the drift\cite{derrida82}.

Another class of systems with ultraslow dynamical properties is
represented by random quantum spin chains at very low
temperatures. For example the asymptotic decay of the zero-temperature
(imaginary time) autocorrelation function
$G(t)=[\langle\sigma_i^x(t)\sigma_i^x\rangle]_{\rm av}$ at the quantum
critical point $(\delta=0)$ is given by\cite{riegerigloi}:
\be
G(t,\delta=0) \sim [\ln t]^{-2x_m}\;,
\label{autocorr0}
\ee
where $x_m$ is the anomalous dimension of the average magnetization.
Away from the critical point, in the Griffiths-McCoy
region\cite{griffiths,mccoy} with $0<\delta\le \delta_G$ the decay
of the autocorrelations is of a power-low form
\be
G(t,\delta) \sim t^{-1/z}\;,
\label{autocorr1}
\ee
where the dynamical exponent $z(\delta) \ge 1$ is a continuous function
of the quantum control-parameter $\delta$. 

Comparing the basic dynamical properties of 
random walks in disordered environments and of random quantum
spin chains one can easily notice close similarities, which hold both
in the critical and off-critical situations. A connection between the
critical properties of directed walks and quantum spin chains is known
for quite some time\cite{igloiturban96}. It was also demonstrated
recently that many surprising properties of the one-dimensional random
transverse-field Ising model (RTIM), a prototype of random quantum
spin chains, can be obtained very simply through random walk arguments
\cite{bigpaper}.

In this Letter we go further and show that behind the similarities
observed before there is a deep connection between the statistical
properties of anomalous diffusion and the critical and off-critical
behavior of the RTIM. We demonstrate this relation by comparing {\it
exact} expressions for the random walk (RW) and that of the RTIM. In
particular we show that the persistence probability of the RW and the
surface magnetization of the RTIM have analogous forms and that the
expressions for the leading eigenvalue of the Focker-Planck (FP)
operator of the RW and the gap of the Hamiltonian of the RTIM are
closely related to each other. We use then this correspondence to
obtain new exact relations for the two systems, among others we
present analytical results about the dynamical exponent $z(\delta)$ in
(\ref{autocorr1}).

We start by considering the one-dimensional random walk with nearest
neighbor hopping, which is characterized by the transition
probabilities $w_{i,i\pm 1}=w(i\to i\pm 1)$ for a random walker to jump from
site $i$ to site $i\pm 1$.
%
%
Here we are particularly interested in the general case, in which the
transition probabilities are not necessarily symmetric\cite{rforce}, i.e. 
$w_{i,i+1}\ne w_{i+1,i}$. 
Moreover, the random walker is confined to a finite number of sites
$i=1,\ldots,L$. At the two ends of this interval, i.e.\ at $i=0$ and
$i=L+1$, we put {\it adsorbing walls}, which is simply modelled by
setting $w_{0,1}=w_{L+1,L}=0$ (i.e.\ the walker cannot jump back into
the system once landed on $0$ or $L+1$).  The time evolution of the
probability distribution of the walk $P_{i,j}(t)$, which is the
probability for the walker to be at time $t$ on site $j$ once started
at time $0$ on site $i$, is fully determined by the Master-equation
\be
\frac{d}{dt}\,\,\underline{P}_i(t)=\underline{\underline{M}}\cdot
\underline{P}_i(t)\;.
\label{master}
\ee
Here
$\underline{P}_i(t)=
\left(P_{i,0}(t),P_{i,1}(t),\ldots,P_{i,L}(t),P_{i,L+1}(t)\right)^T$
and the transition matrix or the Focker-Planck operator
is $(\underline{\underline{M}})_{i,j}=w_{i,j}$
for $i\ne j$ and $(\underline{\underline{M}})_{i,i}=-\sum_j w_{i,j}$
while the initial condition is $P_{i,j}(0)=\delta_{i,j}$. 
The eigenvalue problem of the FP operator in (\ref{master}) is defined by
\be
\underline{\underline{M}}\, \underline{v}_k=\lambda_k \underline{v}_k
~~~,~~~
\underline{u}_k^T\,\underline{\underline{M}} = \underline{u}_k^T 
\lambda_k\;,
\label{eigenprobl}
\ee
and all physical properties of the model can be expressed in terms of
the left and right eigenvectors $\underline{u}_k$ and $\underline{v}_k$,
respectively, and the eigenvalues $\lambda_k$. For instance the
probabilities $P_{i,j}(t)$ are given by
\be
P_{i,j}(t)=\sum_k u_k(i) v_k(j) \exp(\lambda_k t)\;.
\label{probability}
\ee
With adsorbing boundaries the two leading eigenvalues
are zero and the corresponding eigenvectors are
$v_1(i)=\delta_{L+1,i}$,
\beqn
u_1(0)&=&0\label{zerovector}\\
u_1(i)&=&u_1(1)\left[1+\sum_{j=1}^{i-1}\prod_{l=1}^j {w_{l,l-1}
\over w_{l,l+1}}, \right],~~i=2,3,\dots,L+1\;,
\nonumber
\eeqn
while for the other zero mode $v_2(i)=v_1(L+1-i)$  and similarly
$u_2(i)=u_1(L+1-i)$. The value of $u_1(1)$ in (\ref{zerovector})
is fixed by the normalization condition $u_1(L+1)=1$. 

We consider first a quantity that gained considerable interest
recently in related models for anomalous diffusion
\cite{derrida,persistence}: The {\it persistence probability} $P_{\rm
pr}(L,t)$, which is the probability that a walker starting at site
$i=1$ does not cross its starting point until time $t$. Working with
adsorbing sites at $i=0$ and $i=L+1$ we have $P_{\rm
pr}(L,t)=P_{1,L+1}(t)$ and its long time limit $p_{\rm
pr}(L)=\lim_{t\to\infty}P_{\rm pr}(L,t)$ is simply given, via
(\ref{probability}),
$p_{\rm pr}(L)=u_1(1)v_1(L+1)$, where we used the fact
that there is no contribution from the second zero mode, since
$v_2(L+1)=0$. Now with eq(\ref{zerovector}) we have the simple {\it
exact} relation:
\be
p_{\rm pr}(L)
=\left(1+\sum_{i=1}^L\prod_{j=1}^i
\frac{w_{j,j-1}}{w_{j,j+1}}\right)^{-1}
\label{persl}
\ee
Note that $p_{\rm pr}(L)$ is the total fraction of walkers adsorbed by
the right wall ($i$=$L+1$) without ever having crossed the starting
point.  

In a {\it homogeneous} medium with $w_{i,i+1}=w_{i+1,i}=const$ we have
$p_{\rm pr}^{hom}=(L+1)^{-1}$, whereas in an {\it inhomogeneous}
environment with {\it symmetric} transition probabilities
$w_{i,i+1}=w_{i+1,i}$
\be
p_{\rm pr}^{sym}(L)=\left[1+\sum_{i=1}^L {w_{1,0} \over w_{i,i+1}}\right]^{-1}
\propto {D \over L}\;,
\label{symmetric}
\ee
where $D=\left[L\sum_{i=1}^L w_{i,i+1}^{-1}\right]^{-1}$ is the diffusion
constant\cite{zwanzig}. Thus the average persistence probability
in the random symmetric case scales as $\left[p_{\rm pr}^{sym}\right]_{\rm av}
\propto \left[ D \right]_{av}/L$, similarly to the homogeneous case.
From here on we use $[\dots]_{\rm av}$ to denote average over quenched
disorder.

In the general case, with {\it non-symmetric} transition probabilities
we define the control parameter:
\be
\delta_{\rm RW}={[\ln w_\leftarrow]_{\rm av}-[\ln w_\rightarrow]_{\rm av} 
\over \rm{var}[\ln w_\leftarrow]+\rm{var}[\ln w_\rightarrow]}\;,
\label{deltarw}
\ee
where $w_\rightarrow$ ($w_\leftarrow$) stands for transition
probabilities to the right (left), i.e.\ $w_{i,i+1}$ ($w_{i,i-1}$).
For $\delta_{\rm RW}<0$ there is an average drift of the walk towards
the adsorbing site at $i=L+1$, therefore the persistence will have a
finite value in the large
system limit: $\lim_{L \to \infty}
\left[ p_{\rm pr}(L,\delta_{\rm RW}) \right]_{\rm av}>0$, whereas it
goes to zero for $\delta_{\rm RW} \ge 0$. 

Before we proceed with the analysis of the persistence probability
(\ref{persl}) we derive a similar formula for the largest non-zero
eigenvalue of the FP-operator, the {\it absolute} value of which we
denote by $\lambda_{\rm min}$.  According to the relation in
(\ref{probability}) the time-scale $t_r$ of the diffusion process is
set by $t_r \sim \lambda_{\rm min}^{-1}$. It is technically easier to
estimate $\lambda_{\rm min}$ using mixed boundary conditions, which
will, however, not change the scaling behavior of $\lambda_{\rm min}$:
at $i=0$ we assume an adsorbing wall as before, whereas at $i=L$ we
impose a reflecting boundary by setting formally
$w_{L,L+1}=0$. Now, due to different symmetry of the problem there is
only one zero mode of the FP-operator, and the second smallest
eigenvalue in modulus will be $\lambda_{\rm min}$. To determine
$\lambda_{\rm min}$ we use a perturbational method. First, we express
the eigenvalue problem in eq(\ref{eigenprobl}) as
\be
-\tilde{u}(i)w_{i,i-1}+\tilde{u}(i+1)w_{i,i+1}=-u(i) \lambda_{\rm min},
%
\label{perteq}
\ee
where $w_{L,L+1}=0$ and $\tilde{u}(i)=u(i)-u(i-1)~,i=1,2,\dots,L$ in
terms of the components of the left eigenvector $\underline{u} \equiv
\underline{u}_{min}$ and $u(0)=0$. Then, keeping in mind that we are
interesting in situations when $\lambda_{min}(L)$ is a rapidly vanishing
function of $L$ we neglect the r.h.s. of eq(\ref{perteq}) and
derive an approximate expression for the left eigenvector from the
first $L-1$ equations of (\ref{perteq}). Using this result we
obtain an estimate for $\lambda_{\rm min}$ from the last equation of
(\ref{perteq}):
\be
\lambda_{\rm min} \simeq {\tilde{u}(L) \over u(L)} w_{L,L-1}
={u(1) \over u(L)}w_{L,L-1}\cdot
\prod_{j=1}^{L-1} {w_{j,j-1} \over w_{j,j+1}}\:,
\label{lambdamin1}
\ee
which can be transformed into the final form by noticing that
$u(1)/u(L)=p_{\rm pr}(L)$ in eq(\ref{persl}):
\be
\lambda_{\rm min} \sim 
p_{\rm pr}(L) w_{L,L-1}\cdot
\prod_{j=1}^{L-1} {w_{j,j-1} \over w_{j,j+1}}\:.
\label{lambdamin}
\ee

The scaling properties of $\lambda_{\rm min}$ in (\ref{lambdamin}) as
well as the persistence probability $p_{\rm pr}(L)$ in (\ref{persl})
now can easily be derived by establishing a correspondence of these
quantities with the energy gap and surface magnetization,
respectively, of the random transverse-field Ising model in one
dimension defined by the Hamiltonian:
\be
H=-\sum_{i=1}^{L-1} J_i \sigma^x_i \sigma^x_{i+1} -\sum_{i=1}^L h_i
\sigma^z_i\;.
\label{hamilton}
\ee
Here the $\sigma^x_i,\sigma^z_i$ are Pauli matrices at site $i$ and
the $J_i$ exchange couplings and the $h_i$ transverse fields are
random variables. The RTIM in (\ref{hamilton}) has received much
attention
recently\cite{fisher,youngrieger,profiles,riegerigloi,young,bigpaper} and as
was shown in \onlinecite{bigpaper} there are simple expressions for
the surface magnetization $m_s(L)$ as well as for the gap of the
Hamiltonian $\Delta(L)$ of the RTIM in terms of the couplings and
fields.  Comparing those with our results in eqs(\ref{persl}) and
(\ref{lambdamin}) we can set up the following correspondencies
\be
\begin{array}{rcl}
w_{i,i+1} & \longrightarrow & J_i^2\\
w_{i,i-1} & \longrightarrow & h_i^2\\
\delta_{\rm RW} & \longrightarrow & \delta\\ 
p_{\rm pr}(L) & \longrightarrow & m_s^2(L)\\
\lambda_{\rm min}(L) & \longrightarrow & \Delta(L)\\
\mu(\delta_{\rm RW}) & \longrightarrow & 1/z(\delta)
\end{array}
\label{corresp}
\ee
Consequently similar relations hold for the average quantities, when
the transition probabilities (or equivalently the fields and the
couplings) follow the same random modulation.
In the following we use the correspondencies in (\ref{corresp}) to
derive new results.

i) {\it At the critical point} with $\delta_{\rm RW}=0$, which corresponds to
the Sinai's walk\cite{sinai}, the distribution of the leading
eigenvalues of the FP-operator is very broad, $\lambda_{\rm min}(L)$ scales
according to
\be
\lambda_{\rm min}(L) \sim \exp(-{\rm const}\cdot L^{1/2}),~~~\delta_{RW}=0\;,
\label{lambdadistr}
\ee
similarly to the analogous result for the energy gap of the RTIM
\cite{bigpaper}. Note that this scaling relation (\ref{lambdadistr}),
which is consistent with the known relation between relevant time- and
length-scales\cite{sinai}: $L\sim (\log t)^2$, can be most easily
demonstrated by considering the probabilty distribution $P_L(\ln
\lambda_{\rm min})$, which is then expected to scale like
\be
P_L(\ln \lambda_{\rm min})
\sim L^{-1/2}\tilde{p}(\ln\lambda_{\rm min}/L^{1/2})\;,
\ee
as we confirmed numerically.

ii) The scaling behavior of the persistence probability (for zero
drift $\delta_{\rm RW}=0$) in (\ref{persl}) follows also directly from
the analogous result for the surface magnetization $[m_s(L)]_{av}$ of
the RTIM \cite{bigpaper}. Here we just have to mention that $m_s(L)$
at the critical point is not self-averaging, its average value is
dominated by the {\it rare events}, which are of order $O(1)$. From
this follows that the same rare events determine the average of
$m_s^2(L)$, thus the scaling behavior of $[m_s(L)]_{\rm av}$ and
$[m_s^2(L)]_{\rm av}$ are identical. Then, using the correspondence in
(\ref{corresp}) we have the exact result:
\be
[p_{\rm pr}(L)]_{\rm av}
\propto L^{-\theta},~~~\theta=1/2~~~\delta_{\rm RW}=0\;.
\label{pers0}
\ee

iii) In the non-critical situation with $\delta_{\rm RW}\ne 0$ there
is an average drift of the walk towards the site $i=L+1$ ($i=0$) for
$\delta_{\rm RW}<0$ ($\delta_{\rm RW}>0$). In the latter case,
analogously to the surface magnetization of the RTIM \cite{bigpaper},
the average persistence probability vanishes exponentially for large
system sizes
\be
[p_{\rm pr}(L)]_{\rm av} \sim \exp(-L/\xi),
~~~\xi \sim \delta_{\rm RW}^{-2}~~~\delta_{\rm RW}>0\;.
\label{pers+}
\ee
If the average drift of the walk is towards the adsorbing site at
$i=L+1$, thus $\delta_{\rm RW}<0$, there is a non-vanishing infinite
system size limit of the persistence probability (similarly to the
existence of a finite average surface magnetization of the RTIM
\cite{bigpaper}):
\be
\lim_{L \to \infty} [p_{\rm pr}(L)]_{\rm av} 
\sim (-\delta_{\rm RW})^{\beta_{\rm pr}},~~~
\delta_{\rm RW}<0
\label{pers-}
\ee
with $\beta_{\rm pr}=1$, which is approached via an exponential size
dependence. The corresponding correlation length is again given by
$\xi \sim \left(-\delta_{\rm RW}\right)^{-2}$, similar to the case
$\delta_{\rm RW}>0$ in eq(\ref{pers+}). Thus we can conclude that
correlations defined on persistent walks are characterized by the
average critical exponents
\be
\theta=1/2,~~~\nu=2,~~~\beta_{\rm pr}=1\;,
\label{persexp}
\ee
which satisfy the scaling relation $\beta_{\rm pr}=\theta \nu$.

iv) The time-dependent persistence probability $P_{\rm pr}(L,t)$
introduced above is simply given by $P_{\rm
pr}(L,t)=P_{1,L+1}(t)=\sum_k u_k(1)v_k(L+1)\exp(\lambda_k t)$.  In the
random, asymmetric case one expects the scaling relation
\be
\left[P_{\rm pr}(L,\ln t)\right]_{\rm av}=
b^{-\theta} \left[P_{\rm pr}(L/b, \ln t/b^{1/2})\right]_{\rm av}\;,
\label{scaling}
\ee
when lengths are rescaled by a factor $b>1$ and the relation in
(\ref{sinai}) or (\ref{lambdadistr}) between time- and length-scales
are used. Now taking $b=L$ in the limit $t \to \infty$ we recover the
exact result in (\ref{pers0}), on the other hand with $b=(\ln t)^2$ we
have in the large system limit an ultraslow decay $\lim_{L \to
\infty}[P_{\rm pr}(L,t)]_{\rm av} \sim (\ln t)^{-1}$.

\begin{figure}
\epsfxsize=\columnwidth\epsfbox{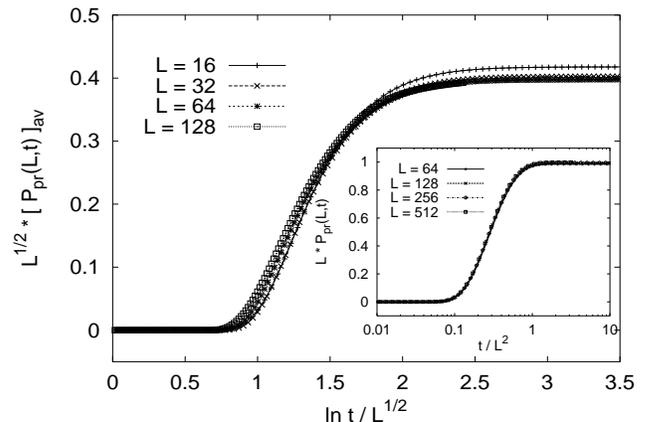}
\caption{
\label{survprob}
Scaling plot of the time-dependent survival probability $P_{\rm
pr}(L,t)$ according to (\protect{\ref{scaling}}) for the asymmetric
hopping model with a uniform distribution of hopping rates (averaged
over 10$^5$ samples). The inset shows the corresponding scaling plot
for the homogeneous case.}
\end{figure}

In the intermediate situation with $\ln t \sim L^{1/2}$ we have (with $b=L$)
the finite size scaling
form
$[ P_{\rm pr}(L,t) ]_{\rm av}\sim L^{-1/2}\overline{p}(\ln t/L^{1/2})$
with $\lim_{y\to\infty}\overline{p}(y)={\rm const}$ --- in contrast to
$P_{\rm pr}^{hom}(L,t)\sim L^{-1}\overline{\overline{p}}(t/L^2)$ in
the homogeneous case.
In Fig. 1 we show corresponding scaling plot
for numerically generated results for finite systems that confirm this
scaling picture.

v) {\it Away from the critical point} we reach the region of
anomalous diffusion, which is equivalent to the Griffiths-McCoy
phase of the RTIM. The relevant energy scale $\lambda_{\rm min}(L)$
($\Delta(L)$ for the RTIM) has a power law scaling behavior:
\be
\lambda_{\rm min}(L)\sim L^{-\mu(\delta_{\rm RW})},~~~
\Delta \sim L^{-1/z(\delta)}\;,
\label{anomscale}
\ee
and according to eq(\ref{corresp}) the two exponents, $\mu$ and $1/z$
correspond to each other. At this point we use the result
that the value of $\mu$ is known exactly from the time dependence of
the average desplacement of the walk in eq(\ref{tmu}) in the
form\cite{derrida82}:
\be
\left[\left({w_{\rightarrow} \over w_{\leftarrow}}\right)^{\mu}\right]
_{\rm av}=1\;.
\label{muexp}
\ee
Essentially this follows from the observation that for any independent
identically distributed random variables $x$ the distribution
$P(\lambda)$ of $\lambda=x_1x_2x_3\cdots$, which is reminiscent of
eq(\ref{lambdamin}), has an algebraic singularity at $\lambda=0$
$P(\lambda)\propto\lambda^{-1+\mu}$ with $\mu$ given by $[x^\mu]_{\rm
av}=1$, see \cite{kesten}.

Consequently we obtain for the dynamical exponent $z$ of the RTIM in
the Griffiths-McCoy phase the implicit equation
\be
\left[\left({J \over h}\right)^{1/z}\right]_{\rm av}=1\;.
\label{zexp}
\ee
Note that for any distribution of $J$ and $h$ one obtains immediately
the result $1/z=2\delta+{\cal O}(\delta^2)$, with $\delta\ll1$ as in
(\ref{deltarw}), as has been observed earlier
\cite{fisher,youngrieger,bigpaper}. However, the exact result
(\ref{zexp}) is not restricted to $\delta\ll1$, but is valid in the
whole Griffiths-McCoy region. For example for the uniform distribution
$$
\pi(J)=\Theta(1-J)\Theta(J);~~~\rho(h)=h_0^{-1}\Theta(h_0-h)\Theta(h)\;,
\nonumber
$$
the dynamical exponent is given by the solution of the equation
\be
z\log(1-z^{-2})=-\ln h_0\quad(=-2\delta)\;,
\label{uniform}
\ee 
The relation (\ref{uniform}) is indeed satisfied by the numerical
estimates for $z$ reported in \cite{youngrieger,young,bigpaper}.

To summarize in this letter we have revealed a fundamental relation
between the anomalous diffusion of random walks in disordered
environments and the slow dynamics, at criticality and in the
Griffiths-McCoy region of the random transverse Ising chain.  With
this analogy at hand we were able to derive a number of new exact
results for both systems.
Many new applications of the above mentioned analogy are obvious:
there is an enormous number of {\it exact} results for various
quantities of random walks in random one-dimensional environments, and
most probably many of them can be directly transferred to
corresponding quantities of random quantum spin chains near the
quantum critical point. It remains a subject of future research to
study in how far these relations carry over to higher dimensions.

\acknowledgements

This work has been supported by the Hungarian National Research Fund
under grants OTKA TO23642 and OTKA TO25139 and by the Ministery of
Education under grant No FKFP 0765/1997. F.\ I. thanks the HLRZ in KFA
J\"ulich, where part of this work has been completed, for kind
hospitality.  H.\ R.'s work was supported by the Deutsche
Forschungsgemeinschaft (DFG).

\end{document}